\documentclass[letterpaper, 10 pt, conference]{ieeeconf} 

\IEEEoverridecommandlockouts               

\usepackage{graphics} 
\usepackage{epsfig} 
\usepackage{times} 
\usepackage{amsmath} 
\usepackage{amssymb} 
\usepackage{pbox}    

\usepackage{tikz}
\usepackage{pgfplots}
\pgfplotsset{compat=1.17}

\usepackage{subfig}
\usepackage{cite}
\usepackage{bm} 
\usepackage{xfrac}

\usepackage{setspace}
\renewcommand{\baselinestretch}{0.966}
\usepackage{microtype}

\usepackage{textcomp}
\usepackage{hyperref}
\usepackage{lipsum}
\newcommand\copyrighttext{
	\footnotesize \textcopyright 2022 IEEE.  Personal use of this material is permitted.  Permission from IEEE must be obtained for all other uses, in any current or future media, including reprinting/republishing this material for advertising or promotional purposes, creating new collective works, for resale or redistribution to servers or lists, or reuse of any copyrighted component of this work in other works.
	DOI: \href{https://doi.org/10.1109/CDC51059.2022.9992615}{10.1109/CDC51059.2022.9992615}}
\newcommand\copyrightnotice{
	\begin{tikzpicture}[remember picture,overlay]
		\node[anchor=south,yshift=10pt] at (current page.south) {\fbox{\parbox{\dimexpr\textwidth-\fboxsep-\fboxrule\relax}{\copyrighttext}}};
	\end{tikzpicture}%
}

\title{\LARGE \bf Approximate Kalman filtering for large-scale systems\\ with an application to hyperthermia cancer treatments}

\author{S.A.N. Nouwens$^{1}$, B. de Jager$^{1}$, M.M. Paulides$^{2,3}$, W.P.M.H. Heemels$^{1}$
\thanks{This research is supported by KWF Kankerbestrijding and NWO Domain AES, as part of their joint strategic research programme: Technology for Oncology II. The collaboration project is co-funded by the PPP Allowance made available by Health$\sim$Holland, Top Sector Life Sciences \& Health, to stimulate public-private partnerships.}
\thanks{$^{1}$Control Systems Technology, Department of Mechanical Engineering, Eindhoven University of Technology, Eindhoven, The Netherlands
}%
\thanks{$^{2}$Electromagnetics for Care \& Cure, Department of Electrical Engineering, Eindhoven University of Technology, Eindhoven, The Netherlands}%
\thanks{$^{3}$Department of	Radiotherapy, Erasmus University Medical Center Cancer Institute, Rotterdam, The Netherlands}%
}

\begin{document}
\bstctlcite{IEEEexample:BSTcontrol} 

\maketitle
\thispagestyle{empty}
\pagestyle{empty}

\copyrightnotice
\begin{abstract}
Accurate state estimates are required for increasingly complex systems, to enable, for example, feedback control. However, available state estimation schemes are not necessarily real-time feasible for certain large-scale systems. Therefore, we develop in this paper, a real-time feasible state-estimation scheme for a class of large-scale systems that approximates the steady state Kalman filter. In particular, we focus on systems where the state-vector is the result of discretizing the spatial domain, as typically seen in Partial Differential Equations. In such cases, the correlation between states in the state-vector often have an intuitive interpretation on the spatial domain, which can be exploited to obtain a significant reduction in computational complexity, while still providing accurate state estimates. We illustrate these strengths of our method through a hyperthermia cancer treatment case study. The results of the case study show significant improvements in the computation time, while simultaneously obtaining good state estimates, when compared to Ensemble Kalman filters and Kalman filters using reduced-order models.
\end{abstract}

\section{Introduction}\label{sec:introduction}
State estimation for large-scale systems is often characterized by a trade-off between computational complexity of the estimation algorithm and the quality of the resulting state estimate. Indeed many schemes, included the ones in \cite{Evensen2003,Haben2011,Barker2004} strive to approximate the Kalman filter, given its optimality as the minimum variance estimator in a linear stochastic setting \cite{Kalman1960}. However, the computational complexity of the estimator is limited when the state estimates are used in real-time, e.g., for state-based controllers, such as, MPC \cite{Mayne2014}. As a result, the state estimate (and control action) must be computed within the sample time, which restricts the applicability of certain state estimation schemes for large-scale systems. 

In this paper, we, therefore, propose a state estimator for a class of large-scale systems that strives to approximate the steady state Kalman filter, while simultaneously being computationally tractable. In particular, we focus on systems derived from Partial Differential Equations (PDEs). For these systems, the state-vector typically has an intuitive interpretation on the spatial domain. This strong link to the spatial domain will be extensively exploited to improve the computational aspects of our estimator. More specifically, we will exploit the (often strong) spatial correlation between the state elements on the spatial domain.

We will show the strengths of our method using a hyperthermia cancer treatment case study \cite{VanderZee2002,Kroesen2019,Hendrikx2018,VilasBoasRibeiro2021b}, which inspired this work, as we observed the computational complexity of standard Kalman filter-based estimators in this setting. Employing model-based estimators to improve Magnetic Resonance Thermometry (MRT) has been investigated already in literature. For example, in \cite{Roujol2012}, a Kalman filter was used for models of small tumors. In \cite{Hendrikx2018, VilasBoasRibeiro2021b} a Kalman filter using a reduced-order model (ROM) was proposed. However, ROM-based Kalman filters are not easy to design and can suffer from long range spurious correlations \cite{Evensen2003}. In \cite{Deenen2021}, a simple Luenberger observer is used. The full state measurements in \cite{Deenen2021} enable an easy to design diagonal innovation gain. However, this simple design does not exploit the strong spatial correlations in the temperature field. In this paper, we will compare our method to several well-known Kalman filter approximations and the Luenberger observer. Our method strives to accurately estimate states for large models, e.g., with $10^6$ states and a computation time in the order of $1-10$ seconds by bridging the gap between computationally intensive Kalman filter and the simple Luenberger observer. We achieve this by interpreting the steady state Kalman filter as a kernel regularized optimization problem. Additionally, we will experimentally show the strengths of our method on an actual hyperthermia treatment setup. 

The remainder of this paper is structured as follows. In Section~\ref{sec:preliminaries}, we introduce the problem and several well-known state estimators. In Section~\ref{sec:key}, we present our proposed state estimation scheme. In Section~\ref{sec:experiment}, we compare our method against the introduced state estimators using a hyperthermia cancer treatment case-study. Last, in Section~\ref{sec:conclusion}, we provide our conclusions and give an outlook for future work.


\section{Preliminaries}\label{sec:preliminaries}
In this section, we start by introducing the system. Hereafter, we summarize the Ensemble Kalman filter, reduced-order model Kalman filter, and Luenberger observer, as these form benchmarks used to compare our newly proposed estimation scheme to.

\subsection{System class}\label{sec:system_class}
In this paper, we consider the discrete-time linear time-invariant (LTI) system 
\begin{subequations}\label{eq:discrete_ss}
	\begin{align}\label{eq:discrete_ss_a}
		\bm{x}_{k+1} &= \bm{Ax}_{k} + \bm{Bu}_k + \bm{w}_k,\quad &\bm{w}_k\sim\mathcal{N}(\bm{0},\bm{Q}),\\ \label{eq:discrete_ss_b}
		\bm{y}_{k} &= \bm{Cx}_k + \bm{\eta}_k,\quad &\bm{\eta}_k\sim\mathcal{N}(\bm{0},\bm{R}),
	\end{align}	
\end{subequations}
where $\bm{x}_k\in\mathbb{R}^{n_x}$, $\bm{y}_k\in\mathbb{R}^{n_y}$, $\bm{u}_k\in\mathbb{R}^{n_u}$ denote the state, output, and input, respectively, at time $k\in\mathbb{N}$. Additionally, $\bm{w}_k\in\mathbb{R}^{n_x}$ and $\bm{\eta}_k\in\mathbb{R}^{n_y}$ denote the zero mean Gaussian distributed process and measurement noise with covariance $\bm{Q}$ and $\bm{R}$, respectively. Moreover, $\bm{A}$, $\bm{B}$, and $\bm{C}$ are matrices of appropriate dimensions. Recall that we focus on the case where $\bm{x}_k\in\mathbb{R}^{n_x}$ with $n_x\gg1$ and where the elements of $\bm{x}_k$ are the result of a spatial discretization, i.e., the $i$-th element is given by $\bm{x}_k^{(i)} = x_k(\bm{r}_i)$ with $x_k:\mathbb{D}\rightarrow\mathbb{R}$ and $\bm{r}_i\in\mathbb{D}$ for $i=1,\dots,n_x$, where $\mathbb{D}\subset\mathbb{R}^D$ denotes the spatial domain.

Below, we review existing observer schemes. In particular for large-scale systems, the observers in Sections~\ref{sec:ensemble_kalman_filter}, \ref{sec:rom-kf}, and \ref{sec:key} require two assumptions.
\begin{itemize}\vspace{-0.3em}
	\item $\bm{R}^{-1}\bm{v}$ and $\bm{R}^{-\top}\bm{v}$ are computationally cheap.
	\item $\bm{Cv}$ and $\bm{C}^\top\bm{v}$ are computationally cheap.  
\end{itemize}\vspace{-0.3em}
A typical $\bm{R}$, that satisfies the first assumption results from uncorrelated measurement noise, i.e., $\bm{R}$ is diagonal.

\subsection{Ensemble Kalman filter}\label{sec:ensemble_kalman_filter}
The Ensemble Kalman filter (EnKF) \cite{Evensen2003} strives to represent the state estimate covariance by an ensemble of state estimates, i.e., $\text{Var}(\bm{\hat{x}}_k) \approx \bm{\bar{X}}_k\bm{\bar{X}}_k^\top$ with $\bm{\bar{X}}_k:=\frac{1}{\sqrt{N-1}}\Big(\begin{bmatrix}\bm{x}_k^{\langle1\rangle}&\hspace{-0.2cm}\dots\hspace{-0.2cm}&\bm{x}_k^{\langle N\rangle} \end{bmatrix} - \frac{1}{N}\sum_{j=1}^{N}\bm{x}_k^{\langle j\rangle}\Big)$. Here, $\bm{\hat{x}}_k$ denotes the state estimate and $\bm{x}_k^{\langle j\rangle}\in\mathbb{R}^{n_x}$ refers to the $j$-th ensemble member and its evolution is given by
\begin{align}\label{eq:enkf_a}
	\bm{x}_k^{\langle j\rangle} &= \bm{A}\bm{\hat{x}}_{k-1}^{\langle j\rangle} + \bm{Bu}_{k-1} + \bm{w}_k^{\langle j \rangle},
\end{align}
where $\bm{w}_k^{\langle j \rangle}$ is a particular process noise realization. When the ensemble is sufficiently large, the statistical properties of $\bm{\bar{X}}_k\bm{\bar{X}}_k^\top$ approximate the true covariance well.

The EnKF updates its ensemble using
\begin{align}\label{eq:enkf_b}
	&\bm{\hat{x}}_k^{\langle j\rangle} = \bm{x}_k^{\langle j\rangle} + \bm{\bar{X}}_k\bm{\bar{Y}}_k^\top \bm{R}^{-1} \\ \nonumber
	&(\bm{I} - \bm{\bar{Y}}_k(\bm{I}+\bm{\bar{Y}}_k^\top\bm{R}^{-1}\bm{\bar{Y}}_k)^{-1}\bm{\bar{Y}}_k^\top\bm{R}^{-1})
	([\bm{y}_k\ \dots\ \bm{y}_k] - \bm{\bar{Y}}_k),
\end{align}
where $\bm{\bar{Y}}_k = \bm{C}\bm{\bar{X}}_k$. Observe that \eqref{eq:enkf_b} uses the assumptions in Section~\ref{sec:system_class} in order to reduce the computational complexity of the matrix inverse using the Woodbury identity \cite{Hager1989}. To obtain a single state estimate from the ensemble, one can simply compute the mean, i.e., $\bm{\hat{x}}_k = \frac{1}{N}\sum_{j=1}^{N}\bm{\hat{x}}_k^{\langle j\rangle}$. Note that there are many variations on the EnKF tailored to specific systems, see \cite{Evensen2003} and the references therein. 

We summarized the computational complexity of each step for the EnKF in Table~\ref{tb:complexity}. Observe that the computational complexity is dominated by $\mathcal{O}(Nn_x^2)$ for large-scale systems (typically $N\ll n_x$). Indeed, propagating the ensemble using \eqref{eq:enkf_a} proved to be computationally intensive, see Section~\ref{sec:experiment}.

\subsection{Reduced-order model Kalman filter}\label{sec:rom-kf}
Large-scale systems are often well-described by smaller ROMs \cite{Antoulas2005, ROWLEY2005}. To this end, it is possible to rewrite the full-order Kalman filter using the ROM given by
\begin{subequations}
	\begin{align}
		\bm{z}_{k+1} &= \bm{A}_r\bm{z}_k + \bm{B}_r\bm{u}_k + \bm{Vw}_k,\\
		\bm{y}_{k} &= \bm{C}_r\bm{z}_k + \bm{\eta}_k,
	\end{align} 
\end{subequations}
where $\bm{z}_k:=\bm{Vx}_k\in\mathbb{R}^{n_r}$ is the reduced-order state vector with $n_r\ll n_x$ and $\bm{V}$ is the projection matrix that satisfies $\bm{V}^\top\bm{V}=\bm{I}$. The Kalman filter using the ROM is given by
\begin{subequations}\label{eq:rom_kalman}
	\begin{align}\label{eq:rom_kalman_a}
		\bm{\bar{z}}_k &= \bm{A}_r\bm{\hat{z}}_{k-1} + \bm{B}_r\bm{u}_{k-1}, \\ \label{eq:rom_kalman_b}
		\bm{\bar{P}}_k &= \bm{A}_r\bm{P}_{k-1}\bm{A}_r^\top + \bm{V}^\top\bm{Q}\bm{V}, \\ \label{eq:rom_kalman_c}
		\bm{K}_k &= (\bm{C}_r\bm{R}^{-1}\bm{C}_r^\top + \bm{\bar{P}}_k^{-1})^{-1}\bm{C}_r^\top\bm{R}^{-1},\\\label{eq:rom_kalman_d}
		\bm{\hat{z}}_k &= \bm{\bar{z}}_k + \bm{K}_k(\bm{\bar{z}}_k - \bm{y}_k),\\\label{eq:rom_kalman_e}
		\bm{P}_k &= (\bm{I} - \bm{K}_k\bm{C})\bm{\bar{P}}_k.
	\end{align}
\end{subequations}
Here, $\bm{\bar{z}}_k$ and $\bm{\hat{z}}_k$ denote the predicted and estimated state, respectively. Note that we reduce the computational complexity of \eqref{eq:rom_kalman_c} using the Woodbury identity \cite{Hager1989}. Crucially, by projecting $\bm{Q}$ to the reduced-order state space, it is likely that part of the process noise is not well described by the reduced-order Kalman filter (ROM-KF). 

The computational complexity for the ROM-KF is provided in Table~\ref{tb:complexity}. Note that for $n_r\ll n_x$, the computational complexity is small, even when the original system is large.

\subsection{Luenberger observer}\label{sec:luenberger_observer}
Last, we introduce the Luenberger observer 
\begin{subequations}\label{eq:luenberger}
	\begin{align}\label{eq:luenberger_a}
		\bm{\bar{x}}_k &= \bm{A\hat{x}}_{k-1} + \bm{Bu}_{k-1},\\\label{eq:luenberger_b}
		\bm{\hat{x}}_k &= \bm{\bar{x}}_k + \bm{K}_\text{Luenberger}(\bm{C\bar{x}}_k - \bm{y}_k),
	\end{align}
\end{subequations}
where $\bm{K}_\text{Luenberger}$ is the user-designed innovation gain. 

For large-scale systems with full state measurements, i.e., $\bm{C}=\bm{I}$, as seen in some hyperthermia cancer treatments \cite{Hensley2015,Deenen2021}, an easy-to-design Luenberger observer uses a diagonal innovation gain, i.e., $\bm{K}_\text{Luenberger}$ is diagonal. We summarized the computational complexity of \eqref{eq:luenberger} when $\bm{C}=\bm{I}$ and $\bm{K}_\text{Luenberger}$ is diagonal in Table~\ref{tb:complexity}. Observe that the computational complexity is dominated by propagating the state forward in time when $\bm{K}_\text{Luenberger}$ is sufficiently sparse.

\begin{table}[!ht]
	\centering
	\caption{Computational complexity of the observers in Sections~\ref{sec:ensemble_kalman_filter} to \ref{sec:luenberger_observer}}
	\scalebox{0.8}{\begin{tabular}{l||l| l| l| l| l}
			Equation  & \eqref{eq:enkf_a}   & \eqref{eq:enkf_b}  & \eqref{eq:rom_kalman_a}  & \eqref{eq:rom_kalman_b}  & \eqref{eq:rom_kalman_c}   \\ \hline
			Complexity & $\mathcal{O}(Nn_x^2)$  & $\mathcal{O}(N^3)$ & $\mathcal{O}(n_r^2)$& $\mathcal{O}(n_r^3)$&  $\mathcal{O}(n_r^3)$
		\end{tabular}}\\ \vspace{3pt}
	\hspace{-5.6mm}\scalebox{0.8}{\begin{tabular}{l||l| l| l| l}
			Equation  & \eqref{eq:rom_kalman_d}   & \eqref{eq:rom_kalman_e}  & \eqref{eq:luenberger_a}  & \eqref{eq:luenberger_b}    \\ \hline
			Complexity & $\mathcal{O}(n_rn_y)$  & $\mathcal{O}(n_r^2n_y)$ & $\mathcal{O}(n_x^2)$& $\mathcal{O}(n_x)$
	\end{tabular}}
	\label{tb:complexity}
	\vspace{-0.5em}
\end{table}
\section{Approximate Kalman filtering for large-scale systems}\label{sec:key}
In this section, we present our proposed approximate Kalman filter that is computationally efficient, intuitive to design, and can provide accurate state estimates, which we will refer to as the least-squares kernel Kalman filter (LSK-KF). To this end, we first introduce an equivalent least-squares form of the Kalman filter \cite{Nash1974}. Hereafter, we present our key insights that enable the efficient computation of the least-squares minimization problem for large-scale systems. 

\subsection{Least-squares Kalman filter}
The least-squares minimization problem that is equivalent to the Kalman filter is
\begin{subequations}\label{eq:lsq-kalman}
	\begin{align}\label{eq:lsq-kalman_a}
		\bm{\bar{x}}_k = & \bm{A\hat{x}}_{k-1} + \bm{Bu}_{k-1},\\ \label{eq:lsq-kalman_b}
		\bm{\hat{x}}_k =& \bm{\bar{x}}_k + \bm{d}_k,\\ \label{eq:lsq-kalman_c}
		\bm{d}_k :=& \underset{\bm{d}}{\arg\min}\ \|\bm{d}\|^2_{\bm{P}_k^{-1}} + \|\bm{y}_k-\bm{C\bar{x}}_k-\bm{Cd}\|^2_{\bm{R}^{-1}} 
	\end{align}
\end{subequations}
with $\|\bm{v}\|_{\bm{M}}^2:=\bm{v^\top Mv}$. Note that \eqref{eq:lsq-kalman} describes the update for the state estimate $\bm{\hat{x}}_k$ but not for the state estimate covariance $\bm{P}_k$. As a result, \eqref{eq:lsq-kalman} is often used with a static $\bm{P}_k$, i.e., $\bm{P}_k = \bm{P}$ for all $k\in\mathbb{N}$, as is the case in the steady state Kalman filter, see, e.g., \cite{Barker2004}. Despite a static state estimate covariance, storing and computing products with $\bm{P}$ is not tractable when $n_x\gg 1$. Moreover, $\bm{P}$ is often ill-conditioned. Hence, solving \eqref{eq:lsq-kalman} can be extremely challenging, especially when the estimator has to be used in real-time applications. 

\subsection{Key concept}
To solve the issue that \eqref{eq:lsq-kalman} is computationally challenging, we can decompose $\bm{P}$ as $\bm{P} = \bm{P}^{\frac{1}{2}}\bm{P}^{\frac{1}{2}}$ \cite{Haben2011}. This decomposition facilitates the coordinate change $\bm{d} = \bm{P}^{\frac{1}{2}}\bm{f}$, which can be used to pre-condition \eqref{eq:lsq-kalman}. Note this change of coordinates simultaneously avoids the inversion of $\bm{P}$ and improves the conditioning of the optimization problem, as we will see.

In our work, we generalize this idea to the decomposition $\bm{P} = \bm{L L}^\top$, where $\bm{L}$ can be any matrix for which matrix-vector products $\bm{Lv}$ and $\bm{L}^\top\bm{v}$ can be computed efficiently. Intuitively, directly designing $\bm{P}$ is analogous to kernel regularization methods, where the kernel specifies desired properties of $\bm{d}$ in \eqref{eq:lsq-kalman} \cite{Pillonetto2014}. Hence, we will refer to this method as the \textit{least-squares kernel Kalman filter} (LSK-KF), as it utilizes a kernel to efficiently approximate the least-squares form of the Kalman filter. Substituting the $\bm{d}_k = \bm{Lf}_k$ and $\bm{P} = \bm{L L}^\top$ (with $\bm{P}_k=\bm{P}$) into \eqref{eq:lsq-kalman_b} and \eqref{eq:lsq-kalman_c} yields
\begin{subequations}\label{eq:lsq-kalman-conditioned}
	\begin{align}
		\bm{\hat{x}}_k =& \bm{\bar{x}}_k + \bm{Lf}_k,\\ \label{eq:lsq-kalman-conditioned_c}
		\bm{f}_k :=& \underset{\bm{f}}{\arg\min}\ \|\bm{f}\|_2^2 + \|\bm{y}_k-\bm{C\bar{x}}_k-\bm{CLf}\|^2_{\bm{R}^{-1}}.
	\end{align}
\end{subequations}
As we will see, interpreting \eqref{eq:lsq-kalman-conditioned_c} (and also \eqref{eq:lsq-kalman}) as a kernel-regularized least-squares problem will provide insight into the design of $\bm{LL}^\top$. We compute \eqref{eq:lsq-kalman-conditioned_c} by solving for the first-order optimality conditions, i.e., the gradient of \eqref{eq:lsq-kalman-conditioned_c} with respect to $\bm{f}$ being zero. This yields the equality
\begin{align}\label{eq:lsq-kalman-conditioned-foo}
	\bm{f} + \bm{L}^\top\bm{C}^\top\bm{R}^{-1}\bm{CLf} = \bm{L}^\top\bm{C}^\top\bm{R}^{-1}(\bm{y}_k-\bm{C\bar{x}}_k),
\end{align}
which can be efficiently solved with iterative linear solvers, such as, for example, the Conjugate Gradient (CG) method \cite{Saad2003}. Solving \eqref{eq:lsq-kalman-conditioned-foo} efficiently with iterative linear solvers requires that the matrix-vector products $\bm{R}^{-1}\bm{v}$, $\bm{Cv}$, $\bm{C^\top v}$, $\bm{Lv}$, and $\bm{L^\top v}$ are computationally cheap (for a $\bm{v}$ of appropriate dimension). This observation also follows from the computational complexity of CG, which is $\mathcal{O}(c\sqrt{\kappa})$, where $c$ denotes the complexity of the matrix-vector products with $\bm{f}$ in \eqref{eq:lsq-kalman-conditioned-foo} and $\kappa$ the condition number of \eqref{eq:lsq-kalman-conditioned_c} \cite{Saad2003}.

\subsection{Choosing L}
The insight that \eqref{eq:lsq-kalman} can be efficiently solved by decomposing a static $\bm{P}$ as $\bm{L L}^\top$, does not solve the problem of choosing an appropriate $\bm{L}$. In order for LSK-KF to approximate the steady state Kalman-filter, we identified two crucial properties
\begin{itemize}
	\item $\bm{LL}^\top$ approximates the asymptotic state estimate covariance of the steady state Kalman filter.
	\item Matrix-vector products $\bm{Lv}$ and $\bm{L}^\top\bm{v}$ must be computationally cheap.
\end{itemize}
In order to satisfy these two assumptions, we propose to design $\bm{L}$ as compositions and linear combinations of so-called ``building blocks". A crucial property of a ``building block" is that matrix-vector products are efficient and have an interpretation on the spatial domain, e.g., spatial smoothing. The interpretation on the spatial domain will be used at the end of this section as a design tool to visualize the covariance $\bm{LL}^\top$, and thus the properties of the resulting state estimator.

In Table~\ref{tb:fast_lv}, we provide a selection of building blocks that can be used to construct $\bm{L}$ using linear combinations or compositions. We will start by introducing one building block in detail. Suppose $\bm{v}\in\mathbb{R}^n$ is the result of spatially discretizing $v:\mathbb{D}\rightarrow\mathbb{R}$ on $\{\bm{r}_1,\dots,\bm{r}_n\}\subset\mathbb{D}$, i.e., $\bm{v}^{(i)} := v(\bm{r}_i)$ for $i=1,\dots,n$. Then, we can approximate the convolution $(l*v)(\bm{r}):=\int_{\mathbb{D}} l(\bm{r}-\bm{r}')v(\bm{r}') \text{d}\bm{r}'$, with $l:\mathbb{D}\rightarrow\mathbb{R}$, by the matrix-vector product, $\bm{Lv}$ where $\bm{L}\in\mathbb{R}^{n\times n}$ and $\bm{L}^{(i,j)} = l(\bm{r}_i - \bm{r}_j)$. The key insight is to evaluate $\bm{Lv}$ on the spatial domain using (with some abuse of notation) $(l*v)(\bm{r}) = \mathcal{F}^{-1}(\mathcal{F}(l)\mathcal{F}(v))(\bm{r})$, where $\mathcal{F}$ denotes the Fourier transform. In fact, the computational complexity of the matrix-vector product $\bm{Lv}$ is generally $\mathcal{O}(n^2)$, but when $\bm{Lv}$ can be computed using the Fast Fourier Transform (FFT), it is reduced to $\mathcal{O}(n\log(n))$ \cite{Cooley1965}. Clearly, some matrix-vector products $\bm{Lv}$ allow for efficient implementation without explicitly constructing $\bm{L}$. We will refer to such matrix-vector products, as matrix-free matrix-vector products.

The remaining building blocks are the following. Multiplying two functions on $\mathbb{D}$, which is equivalent to a matrix-vector product with a diagonal matrix, and can be thus computed with element-wise multiplications. The convolution with a separable kernel $\int_{\mathbb{D}}l(\bm{r},\bm{r}')v(\bm{r}')\text{d}\bm{r}'$, with $l(\bm{r}_i,\bm{r}_j) = \prod_{d=1}^Dl_d(\bm{r}^{(d)}_i,\bm{r}^{(d)}_j)$, which is equivalent to the matrix-vector product where $\bm{L}$ is the result of a Kronecker product, i.e., $\bm{L} = \bigotimes_{d=1}^D\bm{L}_d$ with $\bm{L}_d\in\mathbb{R}^{n_d}$ for $d=1,\dots,D$. This matrix-vector product can be efficiently computed using fast tensor products \cite{Fernandes1998}. Observe that all building blocks have an intuitive interpretation on the spatial domain, which helps in designing an appropriate $\bm{L}$, as we will see.
\begin{table}[!ht]
	\centering
	\caption{A selection of matrix-vector products $\bm{Lv}$ and $\bm{Fv}:=\bm{L^\top v}$ with efficient matrix-free implementations.}
	\label{tb:fast_lv}
	\setlength\tabcolsep{0.15cm}
	\scalebox{0.85}{\begin{tabular}{l|l|l}
		Matrix $\bm{L}$  & Interpretation of $\bm{Lv}$ on $\mathbb{D}$                     & Complexity  \\ \hline
		$\bm{L}^{(i,i)}=l(\bm{r}_i)$& $l(\bm{r})v(\bm{r})$             & $\mathcal{O}(n)$\\ 
		$\bm{L}^{(i,j)} = l(\bm{r}_i-\bm{r}_j)$& $l(\bm{r})*v(\bm{r})$         &  $\mathcal{O}(n\log(n))$\\
		$\bm{L} = \bigotimes_{d=1}^D\bm{L}_d$ & $\prod_{d=1}^Dl_d(\bm{r}^{(d)}_i,\bm{r}^{(d)}_j)*v(\bm{r})$ &  $\mathcal{O}(n\sum_{d=1}^{D}n_d)$\\ \hline
	\end{tabular}}\\ \vspace{0.2cm}
	\scalebox{0.85}{\begin{tabular}{l|l|l}
		Matrix $\bm{F}$  & Interpretation of $\bm{Fv}$ on $\mathbb{D}$ & Complexity \\ \hline
		$\bm{F}^{(i,i)}=l(\bm{r}_i)$	& $l(\bm{r})v(\bm{r})$ & $\mathcal{O}(n)$\\ 
		$\bm{F}^{(i,j)} = l(\bm{r}_j-\bm{r}_i)$ & $l(-\bm{r})*v(\bm{r})$ &  $\mathcal{O}(n\log(n))$ \\
		$\bm{F} = \bigotimes_{d=1}^D\bm{L}^\top_d$ & $\prod_{d=1}^Dl_d(\bm{r}^{(d)}_j,\bm{r}^{(d)}_i)*v(\bm{r})$ &  $\mathcal{O}(n\sum_{d=1}^{D}n_d)$\\ \hline
	\end{tabular}}
\vspace{-0.5em}
\end{table}

Besides computational efficient matrix-vector products, we need to design an appropriate $\bm{L}$ such that $\bm{LL}^\top$ approximates the state estimate covariance of the steady state Kalman filter. Clearly, due to the size of $\bm{LL}^\top$, inspecting the matrix itself is prohibitively difficult. Therefore, we propose using the conditional expected value to evaluate $\bm{LL}^\top$. More specifically, when
\begin{align}
	\begin{bmatrix}
		\bm{v}_a \\ \bm{v}_b
	\end{bmatrix} \sim \mathcal{N}\left(\bm{0},\begin{bmatrix}
		\bm{P}_a & \bm{P}_c \\ \bm{P}_c^\top & \bm{P}_b
	\end{bmatrix}\right),
\end{align}
the conditional expected value is given by
\begin{align}\label{eq:conditional_expectation}
\mathbb{E}(\bm{v}_a\mid\bm{v}_b) = \bm{P}_c\bm{P}_b^{-1}\bm{v}_b
\end{align}
and expresses the expected value of $\bm{v}_a$ conditioned on $\bm{v}_b$. First of all, visualizing $\mathbb{E}(\bm{v}_a\mid\bm{v}_b)$ on the spatial domain will show the correlation of $\bm{v}_a$ with $\bm{v}_b$, which is useful as $\bm{LL}^\top$ can be interpreted as a regularization kernel. Hence, visualizing $\mathbb{E}(\bm{v}_a\mid\bm{v}_b)$ will show state distributions that are associated with a low cost, i.e., state distributions that are promoted by \eqref{eq:lsq-kalman-conditioned_c}. As a result, the conditional expected value is a useful design tool in designing an appropriate $\bm{L}$ by, for example, mimicking the steady state Kalman filter from a small-scale model. We will show how to utilize $\mathbb{E}(\bm{v}_a\mid\bm{v}_b)$ in designing $\bm{L}$ using our hyperthermia case study in Section~\ref{sec:experiment}. Last, observe that \eqref{eq:conditional_expectation} is easy to compute with matrix-vector products when $\bm{v}_b\in\mathbb{R}$ by using $	\begin{bmatrix}
	\bm{P}_c^\top & \bm{P}_b
\end{bmatrix}^\top = \bm{L L}^\top\begin{bmatrix}
	0 & 0 &\dots&  1
\end{bmatrix}^\top$.


\section{Experimental results}\label{sec:experiment}
In this section, we demonstrate the proposed LSK-KF observer by comparing the state estimates with the observers introduced in Section~\ref{sec:preliminaries} using a hyperthermia cancer treatment case-study. Hereto, we first introduce the experimental setup used in actual hyperthermia treatments. Hereafter, we define the state observers, and last, the performance of each observer is quantified by means of the computation time and the state estimation error.

\subsection{Experiment setup}
\begin{figure}[!ht]
	\centering
	\captionsetup[subfloat]{farskip=1pt,captionskip=1pt}
	\subfloat{
		\includegraphics[width=2.25cm]{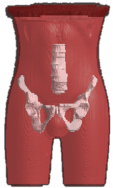}\label{fig:henk_a}
	}
	\subfloat{
		\includegraphics[width=4.5cm]{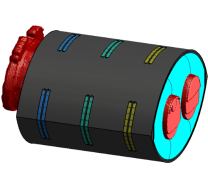}
	}
	\caption{\small Transparent front view of the phantom model (left) and a phantom model placed in the applicator (right).} 
	\label{fig:Henk}
	\vspace{-1em}
\end{figure}

\begin{figure}[!ht]
	\centering
	\captionsetup[subfloat]{farskip=1pt,captionskip=1pt}
	\subfloat[Right focus $b_1(\bm{r})$]{
		\includegraphics[width=3cm]{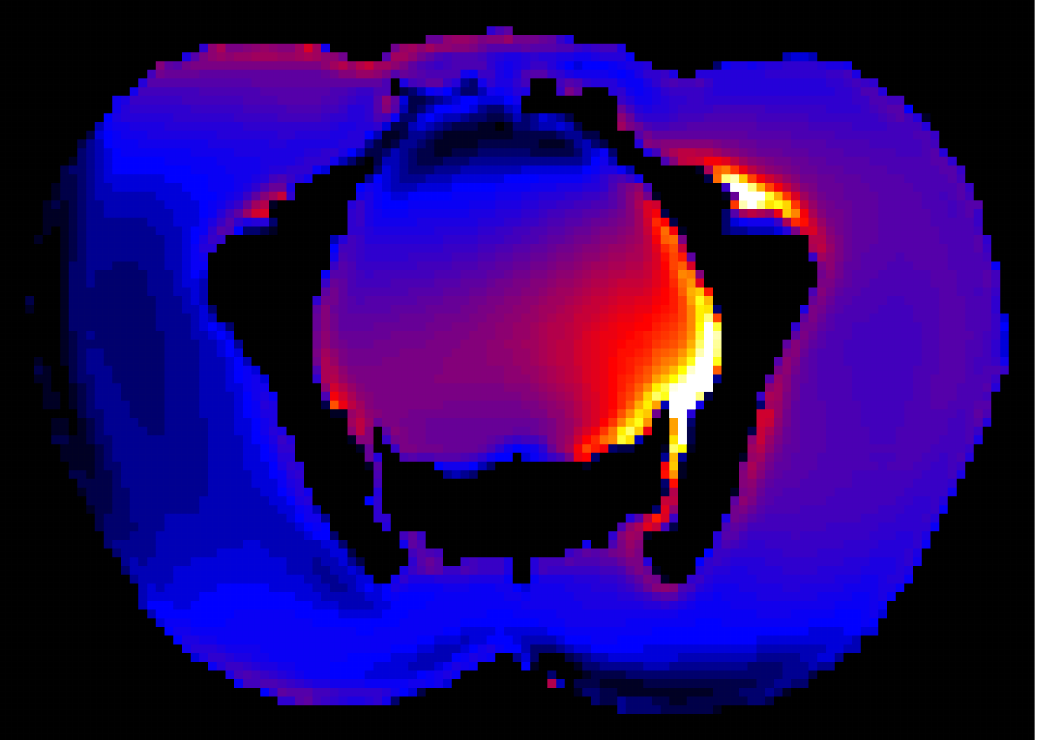}
	}\hspace{0.5cm}
	\subfloat[Left focus $b_2(\bm{r})$]{
		\includegraphics[width=3cm]{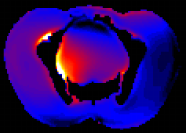}
	}
	\vspace{-0.5em}
	\caption{\small Model-based heat loads $b_1(\bm{r})$ and $b_2(\bm{r})$ on the central transversal slice.}
	\label{fig:sar_focus}
	\vspace{-1em}
\end{figure}

We experimentally validate the observers in Sections~\ref{sec:preliminaries} and \ref{sec:key} using an anthropomorphic phantom of the pelvic region to mimic a typical deep hyperthermia treatment, see \cite{VilasBoasRibeiro2021b}. A voxel-based model of the phantom is shown in Figure~\ref{fig:Henk}. The phantom and hyperthermia applicator can be modeled using a Partial Differential Equation (PDE)
\begin{subequations}\label{eq:pde_henk}
	\begin{align}
		\rho(\bm{r})c(\bm{r})\dot{T}(\bm{r},t) = \nabla\cdot(k(\bm{r})\nabla T(\bm{r},t)) + q(\bm{r},t),
	\end{align}
	for $\bm{r}\in\Omega\subset\mathbb{R}^3$, with $q(\bm{r},t)=b_1(\bm{r})u_1(t)+b_2(\bm{r})u_2(t)$ and convective boundary condition
	\begin{align}
		k(\bm{r})\nabla T(\bm{r},t)\cdot\bm{n} = hT(\bm{r},t),
	\end{align}
	for $\bm{r}\in\partial\Omega$, where $\Omega=\Omega_\text{soft-tissue}\cup\Omega_\text{shell} \cup \Omega_\text{spine} \cup \Omega_\text{pelvis}$ denotes the phantom domain. 
\end{subequations}
Additionally, $T$, $k$, $c$, $\rho$, $b_i$, $u_i$ denote the temperature field with respect to room temperature, thermal conductivity, heat capacity, density, $i$-th heat load, and $i$-th input, respectively. We have two inputs $u_i$ as we consider two RF-antenna inputs, i.e., right focus (corresponding to $b_1(\bm{r})$) and left focus (corresponding to $b_2(\bm{r})$), see Figure~\ref{fig:sar_focus}. The relevant material properties for \eqref{eq:pde_henk} are given in \cite{VilasBoasRibeiro2021b}. Note that we consider temperatures with respect to a steady state, hence, the temperature of the boundary condition is not relevant for this demonstration. 

After discretizing \eqref{eq:pde_henk} in space on a well-chosen nodal grid $\{\bm{r}_1,\dots,\bm{r}_{n_x}\}\subset\Omega$ ($n_x\approx10^6$) and in time with a sample time of $93$ seconds (the time to acquire an MRI scan), we obtain a discrete-time system in the form of \eqref{eq:discrete_ss} with state vector $\bm{x}_k := \begin{bmatrix}
	T(\bm{r}_1,t_k) & \dots & T(\bm{r}_{n_x},t_k)
\end{bmatrix}^\top$ and $t_k = 93k$, $k\in\mathbb{N}$. The output of \eqref{eq:discrete_ss} results from temperature measurements using Magnetic Resonance Imaging (MRI) by exploiting the temperature-dependent Proton Resonance Frequency Shift (PRFS), see \cite{VilasBoasRibeiro2021b} and the references therein. As a result, we define the $i$-th element of the output as $\bm{y}^{(i)}_k:=T(\bm{r}_j,t_k)$ for all $\bm{r}_j\in\Omega_\text{soft-tissue}$. Note that only the soft tissue has a temperature-dependent PRFS. The variance associated with the $i$-th output, i.e., $\bm{R}^{(i,i)}$, is estimated from the measurement data. We refer the interested reader to \cite{VilasBoasRibeiro2021b, Hendrikx2018} for a more in-depth description of the phantom, corresponding thermal model, and the temperature measurement using MRI.

\begin{figure}[!ht]
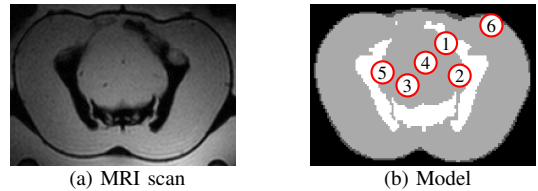

	\centering
	\captionsetup[subfloat]{farskip=1pt,captionskip=1pt}
	\subfloat[MRI scan]{
		\input{probe_loc_mri}
	}\hspace{0.5cm}
	\subfloat[Model]{\label{fig:henk_model}
		\input{probe_loc_model}
	}
	\vspace{-0.5em}
	\caption{\small Temperature probe locations on the central transversal slice of the phantom and model. The white and gray regions in Figure~\ref{fig:henk_model} denote the pelvis and soft-tissue, respectively. }
	\label{fig:probe_loc}
	\vspace{-1.5em}
\end{figure}

To validate the estimated temperature (using any observer), six temperature probes are inserted in the center transversal slice of the phantom using catheter tracks, see Figure~\ref{fig:probe_loc} for the probe locations. These probes serve as a spatially sparse ground truth temperature with a resolution of $0.1^\circ$C and accuracy $\leq0.1^\circ$C. During an experiment of approximately 30 minutes, two sets of inputs are applied to the phantom. The experiment sequence is given by \eqref{eq:experiment_sequence}. Both inputs apply approximately $200$ W to the phantom.
\begin{subequations}\label{eq:experiment_sequence}
	\begin{align}
		u_1(t_k)=&\ 1\ \text{if}\ (2\leq k\leq7),\ \text{and}\ &0,\ \text{otherwise},\\
		u_2(t_k)=&\ 1\ \text{if}\ (8\leq k\leq15),\ \text{and}\ & 0,\ \text{otherwise}.
	\end{align}
\end{subequations}
\vspace{-2em}

\subsection{LSK-KF design}
In this section, we present the design of the LSK-KF observer tailored to the anthropomorphic phantom model \eqref{eq:pde_henk}. For this particular model, we design $\bm{L}$ such that we obtain strong spatial correlations between nearby states within the \textit{same} material and weak correlation between \textit{different} materials. This insight was obtained by solving the discrete-time algebraic Riccati equations to obtain the steady state Kalman filter for a \textit{small-scale} thermal model analogous to \eqref{eq:pde_henk}. Hereto, we choose $\bm{L}$ such that the matrix-vector product $\bm{d} = \bm{Lv}$ with $\bm{d}^{(i)} = d(\bm{r}_i)$ for $i=1,\dots,n_x$ is given by
\begin{subequations}\label{eq:l_design}
	\begin{align}
		d(\bm{r}) &= \sum_{i=1}^{2}\varphi_i(\bm{r})(k(\bm{r})*(\varphi_i(\bm{r})v(\bm{r}))), \\
		\varphi_i(\bm{r}) &= \begin{cases}
			1, & \text{if } \bm{r} \in \Omega_{i}\\
			0, & \text{otherwise}
		\end{cases},\ k(\bm{r}) = \gamma e^{\sigma^{-2}\|\bm{r}\|_2^2}.
	\end{align}
\end{subequations}
Here, $\Omega_{1} = \Omega_\text{soft-tissue}$ and $\Omega_{2} = \Omega_\text{shell} \cup \Omega_\text{spine} \cup \Omega_\text{pelvis}$. Note that for this particular choice $\bm{L}$ in \eqref{eq:l_design}, we have $\bm{Lv} = \bm{L^\top v}$. In \eqref{eq:l_design}, we designed the parameters $\gamma$ and $\sigma$ to match the small-scale steady state Kalman filter by using the conditional expected value \eqref{eq:conditional_expectation}. Observe that the computational complexity of the matrix-vector product \eqref{eq:l_design} scales with $\mathcal{O}(n_x\log(n_x))$, when the convolution with $k(\bm{r})$ is implemented using the FFT \cite{Cooley1965}.

\begin{figure}[!ht]
	\centering
	\captionsetup[subfloat]{farskip=1pt,captionskip=1pt}
	\subfloat[Conditioned on soft-tissue]{
		\includegraphics[width=3.1cm]{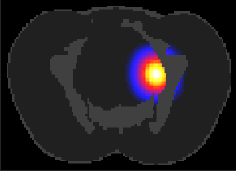}
	}\hspace{0.5cm}
	\subfloat[Conditioned on pelvis]{
		\includegraphics[width=3.1cm]{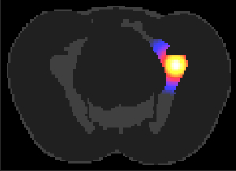}
	}	
	\vspace{-0.5em}
	\caption{\small Conditional expected value conditioned on a state in the soft-tissue (a) and a state in the pelvis (b).}
	\label{fig:cond_exp}
	\vspace{-0.5em}
\end{figure}

In Figure~\ref{fig:cond_exp}, we demonstrate the conditional expected value for two different state elements, one in the soft-tissue material and one in the plastic pelvis. From Figure~\ref{fig:cond_exp}, we conclude that our proposed design \eqref{eq:l_design} indeed achieves strong, localized, spatial correlations within a \textit{single} material and no correlation between \textit{different} materials, as was observed from the steady state Kalman filter using the small-scale model.
\subsection{Observer definitions}
In this section, we briefly present the designs for the observers from Section~\ref{sec:preliminaries}. First, for the EnKF, we compute the process noise realization as $\bm{w} = \bm{Lv}$ with $\bm{v}\sim\mathcal{N}(\bm{0},\bm{I})$ and $\bm{L}$ as in \eqref{eq:l_design}. To evaluated the EnKF we use two ensemble sizes of $N=20$ and $N=100$, respectively. As we will see, the computational tractability of the EnKF will be limited by the ensemble size.

Second, the model \eqref{eq:pde_henk} and corresponding $\bm{V}$ used in the ROM-KF \eqref{eq:rom_kalman_a} is computed using data-driven balanced truncation, and truncating the singular values to include $99.9\%$ of the energy \cite{ROWLEY2005}. The resulting model is of dimension $n_r = 10$. This significant reduction in model order is largely the result of the fact that the model has only two inputs. The reduced-order process noise is estimated by projecting $\bm{LL}^\top$ onto the reduced-order subspace, i.e., $\bm{Q}_r := (\bm{L}\bm{V})^\top(\bm{L}\bm{V})$, where $\bm{LV}$ can be efficiently computed as $\bm{V}$ has few columns.

Last, the diagonal gain for the Luenberger observer is designed as $\bm{K}_\text{Luenberger} := \bm{C}^\top\bm{D}$ with $\bm{D}^{(i,i)} = \frac{d}{d+\bm{R}^{(i,i)}}$. Here, $d\in\mathbb{R}$ is chosen as $d = \tfrac{1}{n_x}\text{trace}(\bm{WW}^\top)$ and $\bm{W}:=\begin{bmatrix}
	\bm{w}^{\langle 1 \rangle} & \dots &\bm{w}^{\langle 500 \rangle}
\end{bmatrix}$, where $\bm{w}^{\langle i \rangle}$ is the same as for the EnKF. As a result, $d$ models the approximate uncertainty of the state estimate for all states.
\subsection{Results}
\begin{figure*}[ht!]
	\centering
	\includegraphics[width=17.7cm]{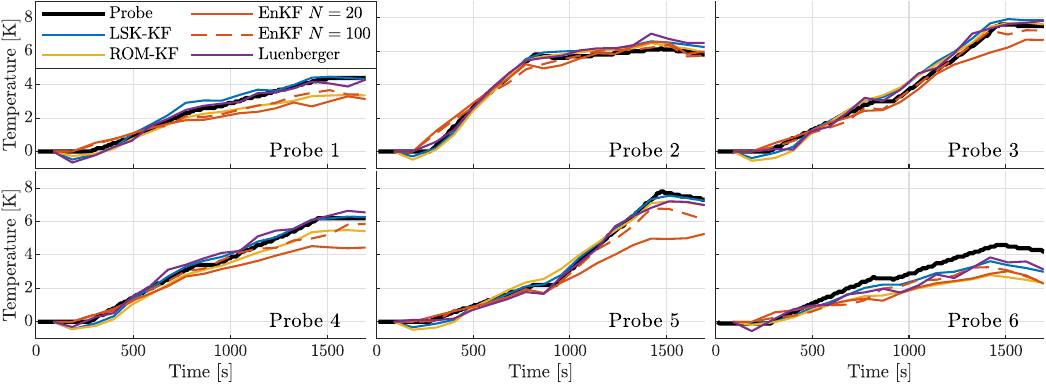}
	\vspace{-1.8em}
	\caption{\small Probe temperature and estimated temperature at the probe locations for different observers.}
	\label{fig:probe_result}
\end{figure*}

\begin{figure}[!ht]
	\centering
	\captionsetup[subfloat]{farskip=1pt,captionskip=1pt,margin = 1em}
	\subfloat[Measurment]{\label{fig:tmaps_a}
		\includegraphics[width=2.7cm]{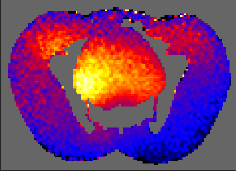}
	}
	\subfloat[LSK-KF]{
		\includegraphics[width=2.7cm]{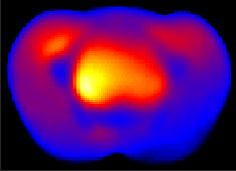}
	} 
	\subfloat[EnKF $N=20$]{
		\includegraphics[width=2.7cm]{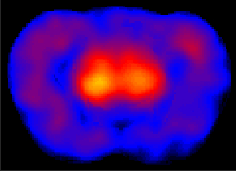}
	} \\
	\subfloat[EnKF $N=100$]{
		\includegraphics[width=2.7cm]{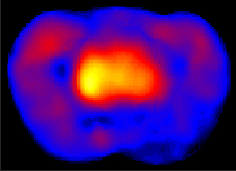}
	} 	
	\subfloat[ROM-KF]{
		\includegraphics[width=2.7cm]{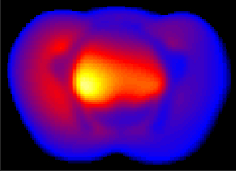}
	}
	\subfloat[Luenberger]{
		\includegraphics[width=2.7cm]{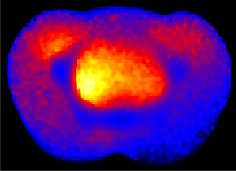}
	}
	\vspace{-0.5em}
	\caption{\small Temperature measurement and estimates at $t=1674$ seconds on the transversal slice centered in the phantom. In Figure~\ref{fig:tmaps_a}, gray areas denote regions with no measurement.}
	\label{fig:tmaps}
	\vspace{-2em}
\end{figure}

Figure~\ref{fig:probe_result} presents the estimated temperature for each observer at the probe locations together with the probe measurements. For a visual comparison between the estimators and the measured temperature, a transversal slice is shown in Figure~\ref{fig:tmaps}. Here, it is clear that that both EnKF estimators have significant temperature irregularities, which are not physically motivated. In the Luenberger observer, the spatially high frequent noise is slightly attenuated, which shows that the spatial correlation between states is not exploited. Both the ROM-KF and the LSK-KF show spatially smooth temperature fields, indicating that the spatial correlation is exploited. 

The results from Figure~\ref{fig:probe_result} are summarized in Table~\ref{tb:rms_probes} using the Root Means Square (RMS) error, the computation time, and the approximate standard deviation of the estimator error for the last two time-steps. The approximate standard deviation of the estimator error assumes the state estimate is given by $\bm{\hat{x}}_k = \bm{x}_k + \bm{e}_k$ where $\bm{x}_k$ is the true state and $\bm{e}_k$ is the estimator error, which is assumed IID. Given this assumption, when $\bm{x}_k \approx \bm{x}_{k+1}$, we can approximate the standard deviation of the elements in $\bm{e}_k$ using
\begin{align}
	\text{STD}(\bm{e}_k) \approx \tfrac{\sqrt{2}}{2}\Big(\tfrac{1}{n_x-1}\sum_{i=1}^{n_x}(\bm{\hat{x}}_k^{(i)} - \bm{\hat{x}}_{k+1}^{(i)} - \mu)^2\Big)^{\sfrac{1\kern-1pt}{2}}
\end{align}
with $\mu = \tfrac{1}{n_x}\sum_{i=1}^{n_x}\bm{\hat{x}}_k^{(i)} - \bm{\hat{x}}_{k+1}^{(i)}$. Note that $\frac{\sqrt{2}}{2}$ corrects the standard deviation as both state estimates $\bm{x}_k$ and $\bm{x}_{k+1}$ have an estimation error.

From Table~\ref{tb:rms_probes}, it is clear that our LSK-KF observer agrees well with the probe data that serve as a ground truth while simultaneously being computationally attractive, even for a model with $10^6$ states. Based on the approximate standard deviation in Table~\ref{tb:rms_probes}, we observe that two consecutive state estimates show little variation when the temperature field is approximately constant between the two time-steps. 

With regard to the other state estimation schemes, the EnKF proved to be computationally intractable, even for modest ensemble sizes. As a result of the modest ensemble size, the state estimates do not agree well with the probes. The ROM-KF, agrees reasonably well with the probes and is computationally cheap. However, we observed that the mode shapes in the projection matrix $\bm{V}$ are not ``rich" enough to describe the required temperature field as there there are significant low frequency spatial correlations in the difference between the measurement and the estimated state. Including more reduced-order states to ROM-KF might solve this problem, but finding relevant modes shapes to extend the ROM-KF is not trivial. The Luenberger observer showed good agreement with the probes, is computationally attractive, but has a higher standard deviation, which is also seen by the spatially high frequent noise in Figure~\ref{fig:tmaps}.

\begin{table}[!ht]
	\centering
	\caption{RMS error of the state estimate relative to the probe measurement, mean computation time, and standard deviation for each observer.}
	\label{tb:rms_probes}
	\setlength\tabcolsep{0.15cm}
\scalebox{0.8}{\begin{tabular}{l||c c c c c}
		     & LSK-KF   & \pbox{20cm}{EnKF \vspace{-1mm} \\  $N=20$ }  & \pbox{20cm}{EnKF \vspace{-1mm}\\ $N=100$}  & ROM-KF  & Luenberger  \\ \hline
		RMS probe 1    & 0.33  & 0.74 & 0.49 & 0.54 & 0.28  \\
		RMS probe 2    & 0.22  & 0.41 & 0.33 & 0.19 & 0.28  \\
		RMS probe 3    & 0.39  & 0.60 & 0.26 & 0.40 & 0.31  \\
		RMS probe 4    & 0.21  & 1.04 & 0.45 & 0.56 & 0.25  \\
		RMS probe 5    & 0.31  & 1.56 & 0.93 & 0.47 & 0.38  \\
		RMS probe 6    & 0.81  & 1.29 & 1.13 & 1.34 & 0.70  \\
		RMS Total& 0.43  & 1.02 & 0.65 & 0.69 & 0.41  \\ \hline
		Time [s]& 7.21  & 40.7 & 200.8& 0.05 & 2.13  \\ \hline
		STD$(\bm{e}_{17})$ & 0.05 & 0.13 & 0.17 & 0.03 & 0.12 \\ \hline
 	\end{tabular}}
\end{table} 

\section{Conclusion and Outlook}\label{sec:conclusion}

In this paper, we presented a computationally attractive state estimation framework for large-scale systems, that is particularly useful for systems derived from partial differential equations. In fact, the proposed LSK-KF can approximate the steady state Kalman filter by approximating the state estimate covariance using computationally attractive building blocks. These building blocks exploit the observation that some matrix-vector products can be efficiently evaluated on the spatial domain. We also proposed the use of the conditional expected value as a computationally efficient tool to design the observer. 

A hyperthermia cancer treatment case study demonstrated the strengths of our estimator. We observed a good correlation to thermal probes, while simultaneously having a low computational complexity. Moreover, we showed how insights obtained from a small-scale model can be used to design the state estimator using the conditional expected value.

This work can be extended in various directions. For instance, expanding the library of building blocks, i.e., matrices with computationally attractive matrix-vector products, is of interest. Additionally, further automating the observer design procedure including stability guarantees for large-scale systems can even further improve the applicability of this work.

\bibliographystyle{IEEEtran}
\bibliography{literature}

\begin{thebibliography}{10}
\providecommand{\url}[1]{#1}
\csname url@rmstyle\endcsname
\providecommand{\newblock}{\relax}
\providecommand{\bibinfo}[2]{#2}
\providecommand\BIBentrySTDinterwordspacing{\spaceskip=0pt\relax}
\providecommand\BIBentryALTinterwordstretchfactor{4}
\providecommand\BIBentryALTinterwordspacing{\spaceskip=\fontdimen2\font plus
\BIBentryALTinterwordstretchfactor\fontdimen3\font minus
  \fontdimen4\font\relax}
\providecommand\BIBforeignlanguage[2]{{%
\expandafter\ifx\csname l@#1\endcsname\relax
\typeout{** WARNING: IEEEtran.bst: No hyphenation pattern has been}%
\typeout{** loaded for the language `#1'. Using the pattern for}%
\typeout{** the default language instead.}%
\else
\language=\csname l@#1\endcsname
\fi
#2}}

\bibitem{Evensen2003}
G.~Evensen, ``{The Ensemble Kalman Filter: theoretical formulation and
  practical implementation},'' \emph{Ocean Dyn.}, pp. 343--367, 2003.

\bibitem{Haben2011}
S.~A. Haben, A.~Lawless, and N.~Nichols, ``{Conditioning of incremental
  variational data assimilation, with application to the Met Office system},''
  \emph{Tellus A Dyn. Meteorol. Oceanogr.}, pp. 782--792, 2011.

\bibitem{Barker2004}
D.~M. Barker, W.~Huang, Y.~R. Guo, A.~J. Bourgeois, and Q.~N. Xiao, ``{A
  three-dimensional variational data assimilation system for MM5:
  Implementation and initial results},'' \emph{Mon. Weather Rev.}, pp.
  897--914, 2004.

\bibitem{Kalman1960}
R.~E. Kalman, ``{A New Approach to Linear Filtering and Prediction Problems},''
  \emph{J. Basic Eng.}, pp. 35--45, 1960.

\bibitem{Mayne2014}
D.~Q. Mayne, ``{Model predictive control: Recent developments and future
  promise},'' \emph{Automatica}, 2014.

\bibitem{VanderZee2002}
J.~van~der Zee, ``{Heating the patient: A promising approach?}'' \emph{Ann.
  Oncol.}, pp. 1173--1184, 2002.

\bibitem{Kroesen2019}
M.~Kroesen, \emph{et~al.}, ``{Confirmation of thermal dose as a predictor of
  local control in cervical carcinoma patients treated with state-of-the-art
  radiation therapy and hyperthermia},'' \emph{Radiother. Oncol.}, pp.
  150--158, 2019.

\bibitem{Hendrikx2018}
R.~Hendrikx, \emph{et~al.}, ``{POD-Based Recursive Temperature Estimation for
  MR-Guided RF Hyperthermia Cancer Treatment: A Pilot Study},'' in \emph{Conf.
  Decis. Control}, 2018, pp. 5201--5208.

\bibitem{VilasBoasRibeiro2021b}
I.~VilasBoas-Ribeiro, \emph{et~al.}, ``{POD-Kalman filtering for improving
  non-invasive 3D temperature monitoring in MR-guided hyperthermia},''
  \emph{Med Phys.}, pp. 1--16, 2022.

\bibitem{Roujol2012}
S.~Roujol, B.~D. de~Senneville, S.~Hey, C.~Moonen, and M.~Ries, ``{Robust
  Adaptive Extended Kalman Filtering for Real Time MR-Thermometry Guided HIFU
  Interventions},'' \emph{IEEE Trans. Med. Imaging}, pp. 533--542, 2012.

\bibitem{Deenen2021}
D.~A. Deenen, \emph{et~al.}, ``{Offset-Free Model Predictive Temperature
  Control for Ultrasound-Based Hyperthermia Cancer Treatments},'' \emph{IEEE
  Trans. Control Syst. Technol.}, pp. 2351--2365, 2021.

\bibitem{Hager1989}
W.~W. Hager, ``{Updating the Inverse of a Matrix},'' \emph{SIAM Rev.}, pp.
  221--239, 1989.

\bibitem{Antoulas2005}
A.~C. Antoulas, \emph{{Approximation of Large-Scale Dynamical Systems}}.\hskip
  1em plus 0.5em minus 0.4em\relax SIAM, 2005.

\bibitem{ROWLEY2005}
C.~W. Rowley, ``{Model Reduction for Fluids, using Balanced Proper Orthogonal
  Decomposition},'' \emph{Int. J. Bifurc. Chaos}, pp. 997--1013, 2005.

\bibitem{Hensley2015}
D.~Hensley, R.~Orendorff, E.~Yu, C.~Danielson, V.~Salgaonkar, and C.~Diederich,
  ``{Model predictive control for treating cancer with ultrasonic heating},''
  in \emph{Am. Control Conf.}, 2015, pp. 220--225.

\bibitem{Nash1974}
R.~Nash, A.~Gelb, and J.~Kasper, \emph{{Applied Optimal Estimation}},
  4th~ed.\hskip 1em plus 0.5em minus 0.4em\relax Cambridge (Mass.): MIT press,
  1974.

\bibitem{Pillonetto2014}
G.~Pillonetto, F.~Dinuzzo, T.~Chen, G.~{De Nicolao}, and L.~Ljung, ``{Kernel
  methods in system identification, machine learning and function estimation: A
  survey},'' \emph{Automatica}, pp. 657--682, 2014.

\bibitem{Saad2003}
Y.~Saad, \emph{{Iterative Methods for Sparse Linear Systems}}.\hskip 1em plus
  0.5em minus 0.4em\relax Society for Industrial and Applied Mathematics, 2003.

\bibitem{Cooley1965}
J.~W. Cooley and J.~W. Tukey, ``{An algorithm for the machine calculation of
  complex Fourier series},'' \emph{Math. Comput.}, pp. 297--301, 1965.

\bibitem{Fernandes1998}
P.~Fernandes, B.~Plateau, and W.~J. Stewart, ``{Efficient descriptor-vector
  multiplications in stochastic automata networks},'' \emph{J. ACM}, pp.
  381--414, 1998.

\end{thebibliography}

\end{document}